\documentclass[a4paper,11pt,psfig]{article}
\usepackage{graphicx}

\newcommand{\sect}[1]{\setcounter{equation}{0}\section{#1}}

\begin{document}
\topmargin 0pt \oddsidemargin 0mm

\renewcommand{\thefootnote}{\fnsymbol{footnote}}
\begin{titlepage}
\begin{flushright}
 hep-th/0411025
\end{flushright}

\vspace{2mm}
\begin{center}
{\Large \bf  Cosmology with Interaction between Phantom Dark
   Energy and Dark Matter and the Coincidence Problem}

\vspace{15mm}
 {\large Rong-Gen Cai\footnote{e-mail address:
cairg@itp.ac.cn}}

\vspace{5mm} {\em CASPER, Department of Physics, Baylor
University, Waco, TX76798-7316, USA \\
Institute of Theoretical Physics, Chinese
Academy of Sciences,\\
 P.O. Box 2735, Beijing 100080, China }

 \vspace{8mm}
 {\large Anzhong Wang\footnote{e-mail address:
 anzhong\_wang@baylor.edu}}

 \vspace{3mm}
 {\em CASPER, Department of Physics, Baylor University, Waco,
 TX76798-7316, USA}

\end{center}

\vspace{20mm} \centerline{{\bf{Abstract}}}
 \vspace{10mm}
We study a cosmological model in which phantom dark energy is
coupled to dark matter by phenomenologically introducing a coupled
 term to the equations of motion of dark energy and dark matter.
This term is parameterized by a dimensionless coupling function
$\delta$,  Hubble parameter and the energy density of dark matter,
and it describes an energy flow between the dark energy and dark
matter. We discuss two cases: one is the case where the
equation-of-state $\omega_e$ of the dark energy is a constant; the
other is that the dimensionless coupling function $\delta$ is a
constant. We investigate the effect of the interaction on the
evolution of the universe, the total lifetime of the universe, and
the ratio of the period when the universe is in the coincidence
state to its total lifetime. It turns out that the interaction
will produce significant deviation from the case without the
interaction.

\end{titlepage}

\newpage
\renewcommand{\thefootnote}{\arabic{footnote}}
\setcounter{footnote}{0} \setcounter{page}{2}

\sect{Introduction}
 A lot of evidence of astronomical observations indicates
  that our universe is currently in accelerated expansion.
  This is first revealed by observing high red-shift supernova Ia~\cite{snae}. Cross
 checks confirm this from the cosmic microwave background
 radiation~\cite{wmap} and large scale structure~\cite{sdss}.
 To explain this accelerated expansion, some proposals have been
 suggested recently in the literature, for instance, by modifying
  Einstein's general relativity in the cosmic distance scale,
  employing  brane world scenario and so on.

 In Einstein's general relativity, however, in order to give an
 explanation for the accelerated expansion, one has to introduce an
 energy component to the energy density of the universe with a large
 negative pressure, which drives the universe to accelerated expand.
 This energy component
 is dubbed dark energy in the literature. All astronomical
 observations indicate that our universe is flat and it consists
 of approximately $72\% $ dark energy, $21\%$ dark matter, $4.5\%$
 baryon matter and $0.5\%$ others like radiation, etc. A simple candidate of the
 dark energy is a tiny positive cosmological constant, which was
 introduced by Einstein in 1917, two years later since he established general
 relativity. If the dark energy is the cosmological constant, one has to answer
 the question why the
 cosmological constant is so small, $ \sim 10^{-122}(M_p)^4$, rather
 than $\sim (M_p)^4$, which is expected from local quantum field theory~\cite{Wen}.
 Here $M_p \sim 10^{19}Gev$ is the Planck mass scale. Although the
 small cosmological constant is consistent with all observational data
 so far, recall that a slow roll scalar field can derive the universe
 to accelerated expand in the inflation models,  it is therefore imaginable
  to use a dynamical field to mimic the behavior of the dark
  energy. In particular, it is hoped that one can solve the so-called
  coincidence problem  by employing a dynamical field.
  The model of scalar field(s) acting as the dark energy is
  called  quintessence model~\cite{quit}. Following $k$-inflation
  model~\cite{kinf}, it is also natural to use a field with
  noncanonical kinetic term to explain the currently accelerated expansion of
  the universe. Such models are referred to as $k$-essence models with
  some interesting features~\cite{kess}. Suppose that the dark
  energy has the equation of state, $p_e=\omega_e \rho_e$, where
  $p_e$ and $\rho_e$ are pressure and energy density, respectively.
   In order to derive the
  universe to accelerated expand, one has to have $\omega_e
  <-1/3$. Note that for the cosmological constant, $\omega_e =-1$; for the
  quintessence model, $-1< \omega_e <-1/3$; and for the $k$-essence
  model, in general one may has $\omega_e >-1$ or $\omega_e<-1$,
  but it is physically implausible to cross $\omega_e =-1$~\cite{vik}.

  It is well known that if $\omega_e <-1$, the dark energy will
  violate all energy conditions~\cite{wald}. However, such dark
  energy models~\cite{cald} are still consistent with observation
  data ($-1.46< \omega_e < -0.78$)~\cite{data}. The dark energy
  model with $\omega_e <-1$ is called phantom dark energy model.
  One remarkable feature of the phantom model is that
  the universe will end with a ``big rip" (future singularity).
  That is, for a phantom dominated universe,
  its total lifetime is finite (see also~\cite{others1,others2}). Before the
   death of the universe, the phantom dark
  energy will rip apart all bound structures like the Milky Way, solar system,
  Earth, and ultimately the molecules, atoms, nuclei, and nucleons of which
  we are composed~\cite{doomsday}.

  Usually  it is assumed that the dark energy is coupled to other matter
  fields only through gravity. Since the first principle is still
  not available to discuss the nature of dark energy and dark matter,
  it is therefore conceivable to consider possible interaction between
  the dark energy and  dark matter. Indeed there exist a lot of
  literature on this subject (see for
  example~\cite{denergy1,denergy2,Dala,Amen,Maje} and references therein).
  In this paper, we also consider an interaction model between the
  dark energy and dark matter by phenomenologically introducing
  an interaction term to the equations of motion of dark energy and
  dark matter, which describes
  an energy flow between the dark energy and dark matter. We
  restrict ourselves to the case that the dark energy is a phantom one.
  Constraint from supernova type
  Ia data on such a coupled dark energy model has been
  investigated very recently~\cite{Maje} (see also
  \cite{Dala,Amen}). Here we are interested in how  such an
  interaction between the phantom dark energy and dark matter
  affects the evolution and total lifetime of the universe.

  On the other hand, one important aspect of dark energy problem
  is the so-called coincidence problem. Roughly specking, the
 question is why the energy densities of dark energy and dark
  matter are in the same order just now. In other words, we live
  in a very special epoch when the dark energy and dark matter
  densities are comparable. Most recently, developing the idea
  proposed by McInnes~\cite{McI}, Scherrer~\cite{Sch} has attacked this
  coincidence problem for a phantom dominated universe. Since the
  total lifetime of the phantom universe is finite, it is therefore possible
  to calculate the fraction of its total lifetime of the universe
  for which the dark energy and dark matter densities are roughly comparable.
  It has been found
  that the coincidence problem can be significantly ameliorated in
  such a phantom dominated universe in the sense that the fraction of the
  total lifetime is not negligibly small.
  In this paper we will also study the effect of the interaction on the
   ratio of the period to its total
  lifetime when the universe is in the coincidence state.

  The organization of this paper is as follows. In the next
  section, we first introduce the coupled dark energy model.
  In Sec.~3 we discuss the case where the phantom dark energy has a
  constant equation-of-state $\omega_e$. In Sec.~4 we study the
  case with a constant coupling function $\delta$ introduced in
  Sec.~2. In this case, the equation-of-state $\omega_e$ will no
  longer be a constant. The conclusion and discussion are presented
   in Sec.~5.

\sect{Interacting phantom dark energy with dark matter}

Let us consider a csomological model which only contains dark
matter and dark energy (generalizing to include the baryon matter
and radiation is straightforward). A phenomenological model of
interaction between the dark matter and dark energy is assumed
through an energy exchange between them. Then the equations of
motion of dark matter and dark energy in a flat FRW metric with a
scale factor $a$ can be written as
\begin{eqnarray}
\label{2eq1}
&& \dot \rho_m +3 H (\rho_m +p_m) = \delta H \rho_m,  \\
\label{2eq2}
 && \dot \rho_e+ 3H(\rho_e+p_e) = -\delta H \rho_m,
\end{eqnarray}
where $\rho_m$ and $p_m$ are the energy density and pressure of
dark matter, while $\rho_e$ and $p_e$ for dark energy, $H \equiv
\dot a/a$ is the Hubble parameter, and $\delta$ is a dimensionless
coupling function. Suppose that the dark matter has $p_m=0$ and
the dark energy has the equation of state $p_e =\omega_e \rho_e$.
Note that in general $\omega_e$ is a function of time, rather than
a constant.  Clearly the total energy density of the universe,
$\rho_t = \rho_m+\rho_e$, obeys the usual continuity equation
\begin{equation}
\label{2eq3}
 \dot \rho_t + 3H (\rho_t +p_t) =0,
 \end{equation}
 with the total pressure $p_t=p_e$. The Friedmann equation is
 \begin{equation}
 \label{2eq4}
 H^2 =\frac{8\pi G}{3} \rho_t,
 \end{equation}
 and the acceleration of scale factor is determined by the equation
 \begin{equation}
 \label{2eq5}
 \frac{\ddot a}{a}=-\frac{4\pi G}{3}\left( \rho_t +3 p_t\right),
 \end{equation}
 where $G$ is the Newton gravitational constant.

 In general the coupling
 function $\delta$ may depend on all degrees of freedom of
 dark matter and dark energy. However, if $\delta$ is dependent
 of the scale factor only, one then can integrate (\ref{2eq1}) and
 obtain
 \begin{equation}
 \label{2eq6}
 \rho_m = \rho_{m,0} a^{-3} e^{\int \delta d\alpha},
 \end{equation}
 where $\alpha = \log a$ and $\rho_{m,0}$ is an integration
 constant.
Substituting this into (\ref{2eq2}), in order to get the relation
between the energy density of the dark energy and scale factor,
one has to first be given the relation of the pressure to energy
density of the dark energy, namely the equation-of-state
$\omega_e$. Here we follow another approach to study the
cosmological model by assuming a relation between the energy
density of dark energy and that of dark matter as
follows~\cite{Maje}:
\begin{equation}
\label{2eq7}
 r \equiv \frac{\rho_e}{\rho_m}=
\frac{\rho_{e,0}}{\rho_{m,0}}\left(\frac{a}{a_0}\right)^{\xi},
\end{equation}
where $\rho_{e,0}$, $a_0$ and $\xi$ are three constants. Set the
current value of the scale factor be one, namely $a_0=1$, then
$\rho_{e,0}$ and $\rho_{m,0}$ can be interpreted as the current
dark energy density and dark matter energy density, respectively.

In this paper we will consider two special cases. One is the case
where $\omega_e$ is kept as  a constant. The other is the case
where the coupling function $\delta$ is a constant.

\sect{Cosmology with a constant equation-of-state of phantom dark
energy}

 In this section we consider the case with a constant $\omega_e$.
  From the relation (\ref{2eq7}) we have
 \begin{equation}
 \label{3eq1}
 \rho_e = \frac{A a^{\xi}}{1+Aa^{\xi}}\rho_t, \  \ \
 \rho_m = \frac{1}{1+A a^{\xi}}\rho_t,
 \end{equation}
 where the constant $A =
 \rho_{e,0}/\rho_{m,0}=\Omega_{e,0}/\Omega_{m,0}$,
 $\Omega_{e,0}$ and $\Omega_{m,0}$ are the fractions of the energy densities
 of dark energy and dark matter at present, respectively. The
 total energy density satisfies
 \begin{equation}
 \label{3eq2}
 \frac{d\rho_t}{da}+\frac{3}{a}\frac{1+(1+\omega_e)A a^{\xi}}{1+A
 a^{\xi}}\rho_t=0.
 \end{equation}
 Integrating this yields
 \begin{equation}
 \label{3eq3}
 \rho_t =\rho_{t,0}a^{-3}[1-\Omega_{e,0}(1-a^{\xi})]^{-3\omega_e/\xi},
 \end{equation}
 where the constant $\rho_{t,0}=\rho_{e,0}+\rho_{m,0}$. Therefore
 the Friedmann equation can be written down as
 \begin{equation}
 \label{3eq4}
 H^2 =H^2_0 a^{-3}[1-\Omega_{e,0}(1-a^{\xi})]^{-3\omega_e/\xi},
 \end{equation}
 with $H_0$ being the present Hubble parameter. By using of (\ref{3eq1}) and
 (\ref{3eq3}), one can get the coupling function $\delta$ from
 (\ref{2eq1}),
 \begin{equation}
 \label{3eq5}
 \delta = 3+\frac{\dot \rho_m}{H \rho_m}=-\frac{(\xi+3\omega_e)A
 a^{\xi}}{1+A a^{\xi}}=-\frac{\xi + 3 \omega_e}{\rho_t}\rho_e,
 \end{equation}
 where an over dot denotes the derivative with respect to the
 cosmic time $t$.
 This can be expressed further as
 \begin{equation}
 \label{3eq6}
 \delta = \frac{\delta_0}{\Omega_{e,0}+(1-\Omega_{e,0})a^{-\xi}},
 \end{equation}
 where $\delta_0 = -\Omega_{e,0}(\xi+3 \omega_e)$.
 We have $\delta (a \to 1) = \delta_0$ and $\delta (a \to \infty)
 = \delta_0/\Omega_{e,0}$.  Therefore we see that when $\xi >-3\omega_e$,
 $\delta <0$, which implies that the energy flow is from the dark matter
 to dark energy.  On the contrary, when $0<\xi <-3\omega_e$,
 the energy flow is from the phantom dark energy to dark matter.
 This can also be understood from the equations of motion
 (\ref{2eq1}) and (\ref{2eq2}).
 Further, we can see from (\ref{3eq5}) that there is no coupling between the
 dark energy and dark matter as $\xi=-3\omega_e$. Of course this
 is true only for case where $\omega_e$ is a constant. In
 addition, we can see from (\ref{3eq1}) and (\ref{3eq3}) that in
 this model, the universe is dominated by the dark matter at early times,
 while dominated by the phantom dark energy at later times.

 The deceleration parameter $q$ is
 \begin{equation}
 \label{3eq7}
  q \equiv -\frac{a\ddot a}{\dot a ^2}=-1 +\frac{\dot H}{H^2}=
  -1+\frac{3}{2}\frac{1-\Omega_{e,0}+(1+\omega_e)\Omega_{e,0}a^{\xi}}
   {1-\Omega_{e,0}(1-a^{\xi})}.
   \end{equation}
Note that $q (a\to \infty) = -1+3(1+\omega_e)/(2\Omega_{e,0})$ and
$q(a\to 1) = -1 +3 \omega_e\Omega_{e,0}/2$, they are always
negative because $\omega_e <0$ and $q(a\to 1) <q (a\to \infty)$.
In Fig.~1-3 we plot the relation of the deceleration parameter to
the red shift defined by $z= 1/a-1$ for the different $\omega_e$
and $\xi$. In plots we take the fraction of the dark energy
$\Omega_{e,0}=0.72$. From figures we can see that for the case
with a fixed $\omega_e$, a larger $\xi$ leads to a smaller red
shift when the universe transits from the deceleration phase to
acceleration phase. On the other hand, for the case with a fixed
$\xi$, a larger $\omega_e$ has a smaller red shift for that
transition from the deceleration to acceleration phase.
\begin{figure}[ht]
\includegraphics[totalheight=1.7in]{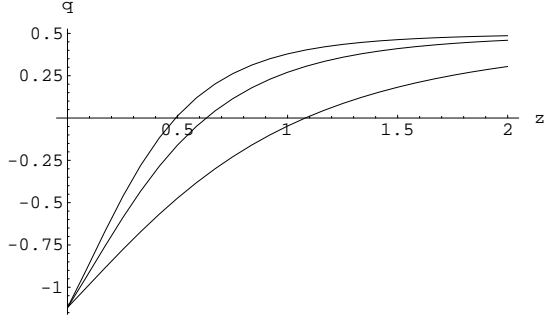}
 \caption{The deceleration parameter $q$ versus the red shift $z$ for the case
 of $\Omega_{e,0}=0.72$ and $\omega_e=-1.5$. Three curves from top to bottom
 correspond to the cases $\xi=5.5$, $4.5$ and $3$, respectively. Note that the
 case of $\xi =4.5 = -3 \omega_e$ is just the case without interaction.}
\end{figure}

\begin{figure}[ht]
\includegraphics[totalheight=1.7in]{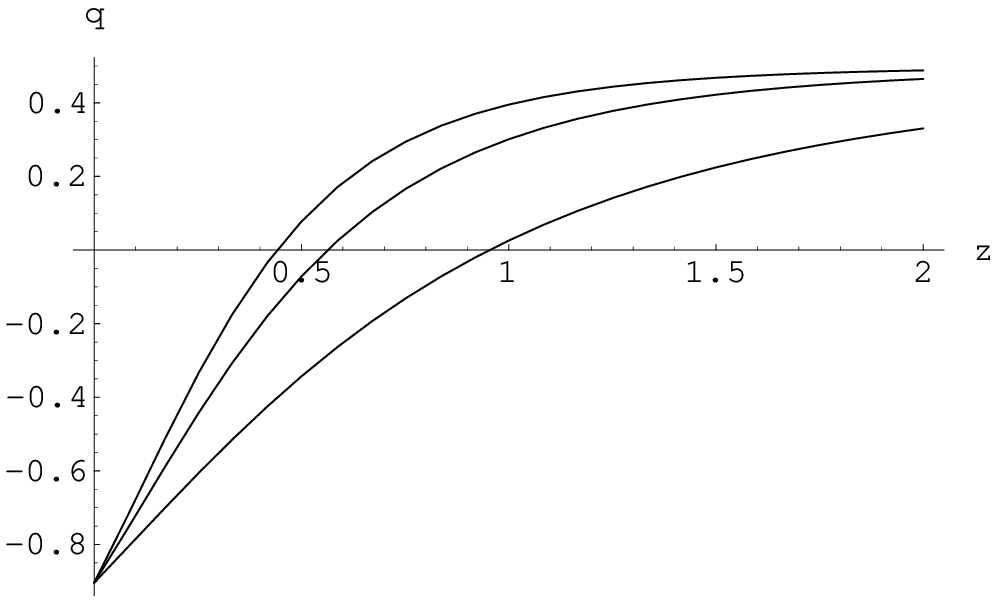}
 \caption{The deceleration parameter $q$ versus the red shift $z$ for the case
 of $\Omega_{e,0}=0.72$ and $\omega_e=-1.3$. Three curves from top to bottom
 correspond to the cases $\xi=5.5$, $4.5$ and $3$, respectively. }
\end{figure}
\begin{figure}[ht]
\includegraphics[totalheight=1.7in]{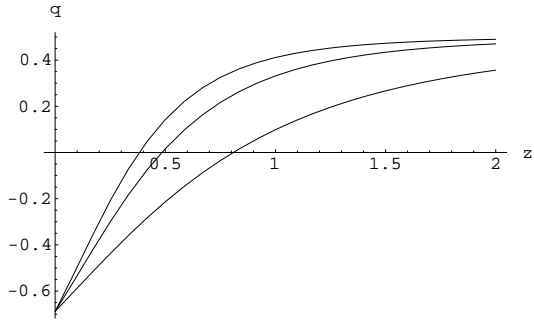}
 \caption{The deceleration parameter $q$ versus the red shift $z$ for the case
 of $\Omega_{e,0}=0.72$ and $\omega_e=-1.1$. Three curves from top to bottom
 correspond to the cases $\xi=5.5$, $4.5$ and $3$, respectively. }
\end{figure}

The total lifetime of the universe can be obtained by integrating
the Friedmann equation (\ref{3eq4}). It is
\begin{equation}
\label{3eq8}
 t_U = H^{-1}_0 \int^{\infty}_0 da\ a^{1/2}
 [1-\Omega_{e,0}(1-a^{\xi})]^{3\omega_e /2\xi}.
 \end{equation}
 Here we are interested in the change of the lifetime due to the
 interaction between the dark energy and dark matter. Note that
 when $\xi =-3\omega_e$, the interaction disappears. Denote the
 total lifetime by $t_T$ for this case, one has
 \begin{equation}
 \label{3eq9}
 t_T=H^{-1}_0 \int^{\infty}_0 da\ a^{1/2}
 [1-\Omega_{e,0}(1-a^{-3\omega_e})]^{- 1/2}.
 \end{equation}
 Denote the ratio of the lifetimes $t_U$ to $t_T$ by $g$:
 \begin{eqnarray}
\label{3eq10}
 g  &\equiv & \frac{t_U}{t_T}=\frac{\int^{\infty}_0 da\ a^{1/2}
 [1-\Omega_{e,0}(1-a^{\xi})]^{3\omega_e /2\xi}}{\int^{\infty}_0 da\ a^{1/2}
 [1-\Omega_{e,0}(1-a^{-3\omega_e})]^{- 1/2}} \nonumber \\
   &=&
   \frac{\int^{\infty}_0\Omega_{e,0}^{-3/2\xi}(1-\Omega_{e,0})^{3(1+\omega_e)/2\xi}
   r^{3/2\xi -1}(1+r)^{3\omega_e/2\xi}dr}
   {\int^{\infty}_0
   \Omega_{e,0}^{1/2\omega_e}(1-\Omega_{e,0})^{-(1+\omega_e)/2\omega_e}
    r^{-1/2\omega_e -1}(1+r)^{-1/2}dr}.
 \end{eqnarray}
\begin{figure}[ht]
\includegraphics[totalheight=1.7in]{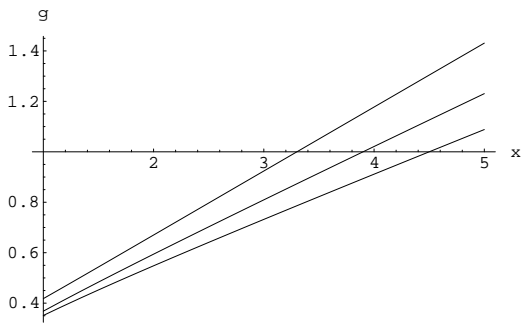}
 \caption{The ratio $g$ of total lifetimes versus the parameter $\xi=x$ for the case
 of $\Omega_{e,0}=0.72$. Three curves from bottom to top
 correspond to the cases $\omega_e=-1.5$, $-1.3$ and $-1.1$, respectively.}
\end{figure}
In Fig.~4 we plot the ratio $g$ for three different
equation-of-states $\omega_e =-1.5$, $-1.3$ and $-1.1$. Clearly,
for a fixed $\omega_e$, the universe with a larger $\xi$ has a
longer lifetime, while for a fixed $\xi$, a larger $\omega_e$
leads to a longer lifetime.  Note that in Fig.~4 the three points
where three curves cross the $\xi$ axis correspond to the
situation ($\xi =-3 \omega_e$) without interaction between the
dark energy and dark matter. In Fig.~5 the ratio $g$ is plotted
versus the parameters $\xi$ and $\omega_e$. Note that for a
constant equation-of-state $\omega_e$, the total lifetime of the
universe approximately is~\cite{Sch}:
$$t_T= \frac{\omega_e}
{1+\omega_e} t_m, $$
 where $t_m$ is the age of the universe when
the matter and phantom dark energy densities are equal. When
$\omega_e = -1.5$, $-1.3$ and $-1.1$, one has $t_T = 3t_m$, $ 4.3
t_m$, and $11 t_m$, respectively. In order to satisfy the
observation data, $\xi$ has to be chosen at least so that $g >
1/3$, $1/4.3$ and $1/11$, respectively. We can see from Fig.~4
that the constraint of the age of the universe on the model is
very weak; the parameter $\xi$ can be as small as $1$. On the
contrary, if one requires that the transition of the universe from
the deceleration phase to acceleration phase happens around at red
shift $z \le 1$,  one can see from Fig.~1-3 that one needs to take
$\xi \ge 3$. Note that the best fit of the $\Lambda CDM$ model
indicates that this transition happens at $z\sim 0.5$. However, it
is allowed that the transition happens at $z\in (0.3 \sim 1)$ in
the different models. Clearly it is quite necessary to make a
detailed numerical analysis and to give constraints on the
parameters of the coupled dark energy model by using of the data
from supernova, cosmic microwave background radiation and large
scale structure. Here let us just mention that a partial analysis
of the model according to the data of supernova has been done more
recently~\cite{Maje}. The results shows that the supernova data
favor a negatively coupled phantom dark energy with $\omega_e <-1$
and $-10 < \delta_0 <-1$.
\begin{figure}[ht]
\includegraphics[totalheight=1.7in]{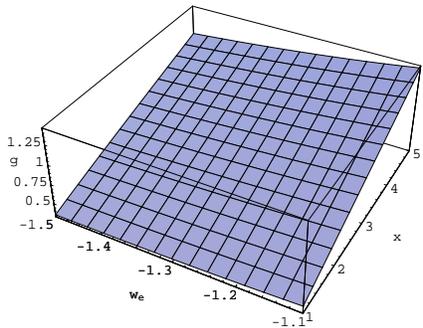}
 \caption{The ratio $g$ of total lifetimes versus the parameters $\xi=x$ and $\omega_e$
  in the case of $\Omega_{e,0}=0.72$. }
\end{figure}

Next we turn to the coincidence problem. Following~\cite{Sch}, we
calculate the ratio of the period when the universe is in the
coincidence state to the total lifetime of the universe. That is,
we will calculate the quantity
\begin{equation}
f = \frac{t_c}{t_U},
\end{equation}
where  $t_c$ is defined by
\begin{equation}
t_c =H^{-1}_0 \int^{a_2}_{a_1} da\ a^{1/2}
 [1-\Omega_{e,0}(1-a^{\xi})]^{3\omega_e /2\xi}.
 \end{equation}
 During the coincidence state, the energy density of dark energy is comparable to
  that of dark matter and the scale factor evolves from $a_1$ to $a_2$.
 What is the exact meaning by the term ``comparable"?
  This is not a well-defined  question in order to determine the scale
 factors $a_1$ and $a_2$. In \cite{Sch}, Scherrer defined a scale
 of the energy density ratio $r_0$ so that the dark energy and
 dark matter densities are regarded as to be comparable if they differ by less
 than the ratio $r_0$ in either direction. He found that the
 ratio varies from $1/3$ to $1/8$ as $\omega_e$ varies from $-1.5$
 to $-1.1$ if $r_0=10$ in a phantom dark energy model without
 interaction between the dark energy and dark matter. In this sense indeed the
 coincidence problem is significantly ameliorated in the phantom model
 because the ratio is not so small. As a result it is not so
 strange that we live in the epoch when the energy densities of
 dark energy and dark matter are of the same order.
 Now we want to see how the fraction varies when the phenomenological
 interaction is introduced.

The fraction of the total lifetime of the universe for which the
universe is in the coincidence state, turns out to be
\begin{eqnarray}
f &= &
  \frac{\int^{r_0}_{1/r_0}
   r^{3/2\xi -1}(1+r)^{3\omega_e/2\xi}dr}
   {\int^{\infty}_0 r^{3/2\xi -1}(1+r)^{3\omega_e/2\xi}dr }
   \nonumber \\
   &=&\frac{\frac{2}{3}(r_0^{3/2\xi} \
   _2F_1[\frac{3}{2\xi},-\frac{3\omega_e}{2\xi},1+\frac{3}{2\xi},-r_0]
   -r_0^{-3/2\xi}
   \ _2F_1[\frac{3}{2\xi},-\frac{3\omega_e}{2\xi},1+\frac{3}{2\xi},
   -\frac{1}{r_0}])}{\frac{\Gamma[3/2\xi]\Gamma[-3(1+\omega_e)/2\xi]}
   {\Gamma[-3\omega_e/2\xi]}}.
\end{eqnarray}
Note that this ratio is independent of the current density
parameter $\Omega_{e,0}$. In Fig.~6 and 7 we plot the ratio $f$
versus the scale $r_0$ for different parameters $\omega_e$ and
$\xi$. Clearly for the case with fixed $\omega_e$ and $\xi$,  a
larger $r_0$ leads to a larger ratio $f$.  On the other hand, for
the case with fixed $r_0$ and $\omega_e$, a smaller $\xi$ gives us
a larger ratio. For example, we see from Fig.~6 that when $r_0=10$
and $\omega_e=-1.5$, the ratio $f \sim 0.45$ for $\xi=3$, more
large than the case ($f=1/3$) without the interaction. Note that
the middle curve in Fig.~6 corresponds to the case without the
interaction ($\xi =-3\omega_e$). In Fig.~8, we plot the ratio $f$
versus the parameters $\xi$ and $r_0$ for a fixed $\omega_e=-1.3$.
From Fig.~6-8, one can see that indeed the period when the
universe is in the coincidence state is comparable to its total
lifetime. In addition, We
 note  from (\ref{2eq7}) that a larger $\xi$ means that the
universe will be dominated more quickly by the phantom dark
energy, it corresponds to have a longer total lifetime of the
universe, which can be seen from Fig.~4 and 5. Thus it is natural
to have a smaller fraction $f$ than the case with a smaller $\xi$.

\begin{figure}[ht]
\includegraphics[totalheight=1.7in]{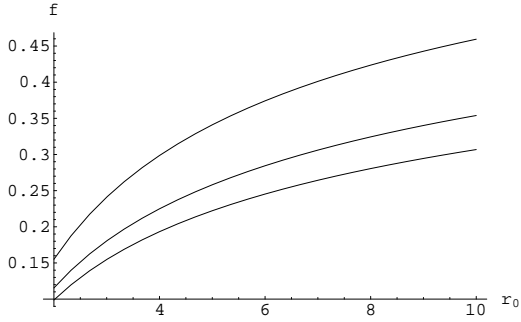}
 \caption{The ratio $f$  versus the parameter $r_0$ for the case of $\omega_e=-1.5$.
 Three curves from top to bottom correspond to the cases of $\xi=3$, $4.5$ and $5.5$,
 respectively.}
\end{figure}
\begin{figure}[ht]
\includegraphics[totalheight=1.7in]{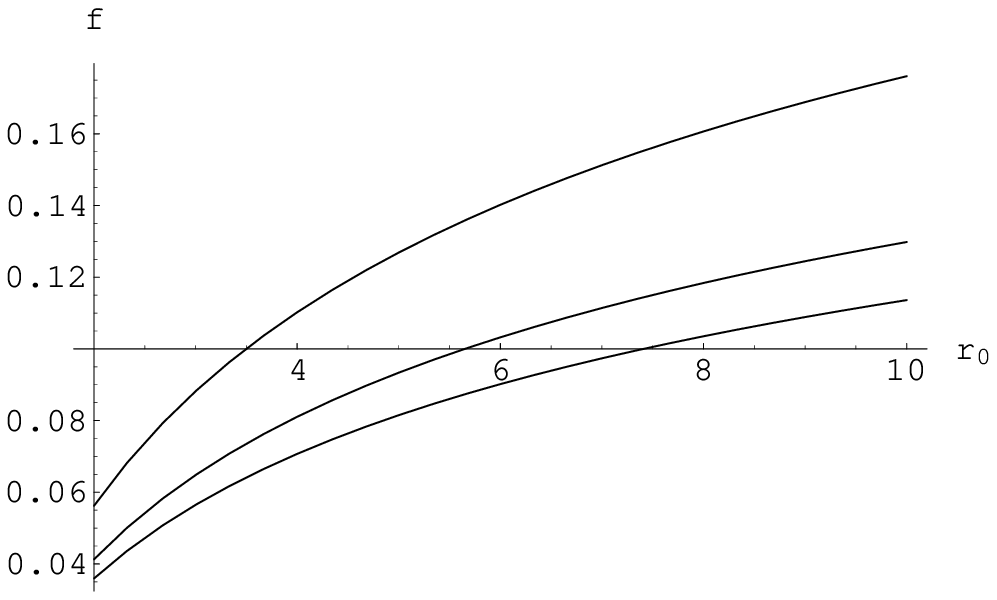}
 \caption{The ratio $f$  versus the parameter $r_0$ for the case of $\omega_e=-1.1$.
 Three curves from top to bottom correspond to the cases of $\xi=2$, $3.3$ and $4$,
 respectively. }
\end{figure}
\begin{figure}[ht]
\includegraphics[totalheight=1.7in]{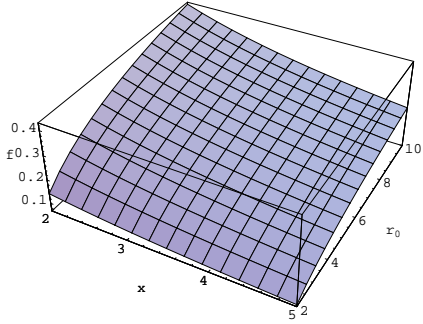}
 \caption{The ratio $f$ versus the parameters $r_0$ and $\xi$ for the case
 of $\omega_e=-1.3$. In the figure $x=\xi$.}
\end{figure}

\sect{Cosmology with a constant coupling parameter}

In this section we consider the case with a constant coupling
function $\delta$. In this case, we have the energy density of
dark matter
\begin{equation}
\label{4eq1}
 \rho_m=\rho_{m,0}a^{-3+\delta}.
 \end{equation}
 And then the dark energy density has the relation to the
 scale factor
 \begin{equation}
 \label{4eq2}
 \rho_e=\rho_{e,0}a^{-3+\delta +\xi}.
 \end{equation}
 The Friedmann equation turns out to be
 \begin{equation}
 \label{4eq3}
 H^2 =H^2_0(\Omega_{m,0}a^{-3+\delta}
 +\Omega_{e,0}a^{-3+\delta+\xi }).
 \end{equation}
 In this case, the equation-of-state $\omega_e$ of the dark energy
 will depend on time (scale factor). From (\ref{2eq2}), we can obtain
 \begin{equation}
 \label{4eq4}
  \omega_e = -\frac{\delta+\xi}{3}
  -\frac{\delta}{3}\frac{\Omega_{m,0}}{\Omega_{e,0}}a^{-\xi}.
  \end{equation}
  When $\delta =0$, one has $\xi =-3\omega_e$. This situation is
  just the case without the interaction.
  From (\ref{4eq2}) and (\ref{4eq4}), one can see that in order the dark energy to
  be  phantom, $\delta+\xi > 3$ has to be satisfied so that the dark energy density
  increases with the scale factor. When $\delta+\xi=3$, although
  the dark energy density keeps as a constant, it does not act as a
  cosmological constant due to the interaction between the dark
  energy and dark matter. We see from (\ref{4eq4}) that
  \begin{equation}
  \omega_{e}(a \to 0) = -sign(\delta)\cdot \infty, \ \ \omega_{e} (a \to 1)=-\frac{\delta+\xi}{3}
  -\frac{\delta}{3}\frac{\Omega_{m,0}}{\Omega_{e,0}}, \ \ \
  \omega_e(a \to \infty) = -\frac{\delta+\xi}{3}.
  \end{equation}
  It is interesting to note that $\omega_e$ diverges as $a \to 0$.
  This is due to the fact that the dark energy density (\ref{4eq2})
  approaches to zero very quickly, while the energy density of dark matter
  (\ref{4eq1}) goes to infinity as $a \to 0$, in order for the
  equation (\ref{2eq2}) to hold, the parameter $\omega_e$ has to
  go to infinity. In fact, the dark energy
  does not play any role at early times, which can be seen
  from the Friedmann equation (\ref{4eq3}). Therefore here the
  divergence of $\omega_e$ does not make any sense in physics.

  The deceleration parameter is found to be
  \begin{equation}
  \label{4eq6}
  q = -1 + \frac{1}{2}\frac{(3-\delta)\Omega_{m,0}+
  (3-\delta-\xi)\Omega_{e,0}a^{\xi}}{\Omega_{m,0}+\Omega_{e,0}a^{\xi}},
  \end{equation}
  which has $q= -1+(3-\delta -\xi \Omega_{e,0})/2$ when $a=1$ and
  $q= (1-\delta-\xi)/2$ when $a\to \infty$. In Fig.~9 and 10 we
  plot the deceleration parameter versus the red shift for the case with
  a fixed $\xi=4$, different $\delta$, and the case with a fixed $\delta=0.3$,
  different $\xi$, respectively. Note that the coupling parameter
  $\delta$ will affect the expansion law (\ref{4eq1}) of dark matter
  density. We do not expect that the usual behavior ($\rho_m \sim
  a^{-3}$) will be changed much. So we take the value of $\delta$ so that the
  exponent is changed within $10\%$.  That is, $\delta $ is taken
  to be in the range of $ (-0.3,0.3)$. Of course, in order to get
  a correct constraint on the parameter $\delta$ from the observation
  data,  a detailed numerical analysis has to be done.
  We see from Fig.~9 that for the case with a
  given $\xi$, a larger $\delta$ leads to  a smaller red shift when
  the universe transits from a deceleration phase to an
  acceleration phase, while Fig.~10 tells us that for the case with a fixed
  $\delta$, a larger $\xi$ gives us a smaller red shift.
\begin{figure}[ht]
\includegraphics[totalheight=1.7in]{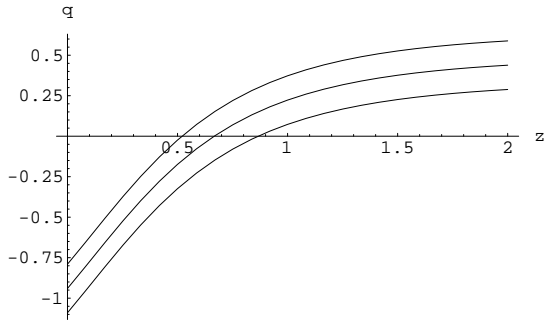}
 \caption{The deceleration parameter $q$ versus the red shift $z$ for the case
 of $\Omega_{e,0}=0.72$ and  $\xi=4$.  Three curves from top to bottom
 correspond to the cases $\delta=0.3$, $0$ and $-0.3$, respectively. }
\end{figure}
\begin{figure}[ht]
\includegraphics[totalheight=1.7in]{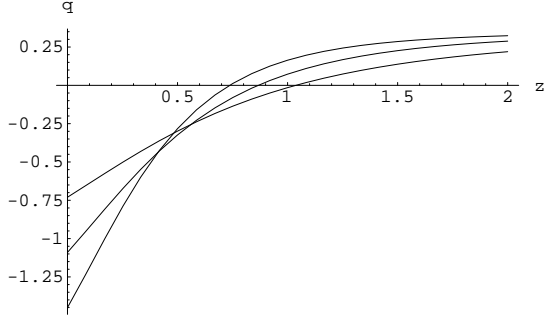}
 \caption{The deceleration parameter $q$ versus the red shift $z$ for the case
 of $\Omega_{e,0}=0.72$ and  $\delta=0.3$. Three curves from bottom to
 top at the $q$ axis correspond to the cases $\xi=5$, $4$ and $3$, respectively. }
\end{figure}

  From (\ref{4eq3}) we can get the total lifetime of the universe
  \begin{equation}
  \label{4eq7}
  t_U = H_0^{-1}\int^{\infty}_0 da \ a^{-1}
  (\Omega_{m,0}a^{-3+\delta}+\Omega_{e,0}a^{-3+\delta+\xi})^{-1/2}.
  \end{equation}
  We now consider the effect of the interaction on the total
  lifetime. Note that the total lifetime of the
  universe without the interaction is
  \begin{equation}
  \label{4eq8}
t_T = H_0^{-1}\int^{\infty}_0 da \ a^{-1}
  (\Omega_{m,0}a^{-3}+\Omega_{e,0}a^{-3+\xi})^{-1/2}.
  \end{equation}
  Denote the ratio $t_U/t_T$ by $g$,  we can express this
  as
  \begin{equation}
  g = \frac{\int^{\infty}_0 dr
  (\frac{1-\Omega_{e,0}}{\Omega_{e,0}})^{(3-\delta)/2\xi}
     r^{(3-\delta-2\xi)/2\xi}(1+r)^{-1/2}}{\int^{\infty}_0 dr
  (\frac{1-\Omega_{e,0}}{\Omega_{e,0}})^{3/2\xi}
     r^{(3-2\xi)/2\xi}(1+r)^{-1/2}}.
  \end{equation}
In Fig.~11, the ratio $g$ is plotted versus the parameters $\xi$
and $\delta$ for the case $\Omega_{e,0}=0.72$. We see that the
case $\delta >0$ is quite different from the case of $\delta<0$.
For the case with a fixed $\delta>0$, a larger $\xi$ leads to a
longer lifetime of the universe. On the contrary, for the case
with a fixed $\delta <0$, a smaller $\xi$ gives us a longer
lifetime. Further, for the case with a fixed $\xi$, a smaller
$\delta$ has a longer lifetime of the universe.
\begin{figure}[ht]
\includegraphics[totalheight=1.7in]{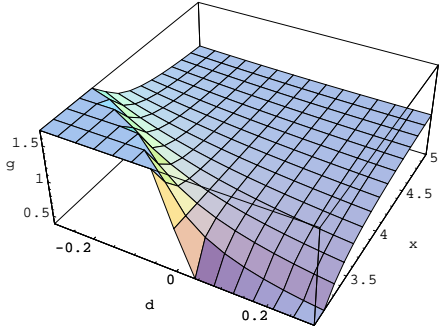}
 \caption{The ratio $g$ of total lifetimes versus the parameters $\xi=x$ and $\delta=d$
 for the case of $\Omega_{e,0}=0.72$.}
\end{figure}

  Finally we consider the fraction of the total lifetime for which the universe is
  in the coincidence state.  As the case with a constant $\omega_e$
  considered in the previous section, we calculate the following
  ratio:
  \begin{equation}
  f = \frac{\int^{r_0}_{1/r_0} dr \
     r^{(3-\delta-2\xi)/2\xi}(1+r)^{-1/2}}
     {\int^{\infty}_{0} dr\
     r^{(3-\delta-2\xi)/2\xi}(1+r)^{-1/2}}.
     \end{equation}
\begin{figure}[ht]
\includegraphics[totalheight=1.7in]{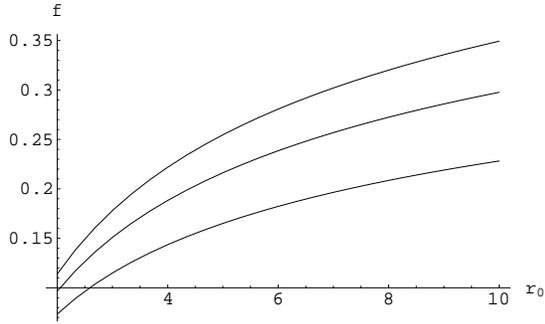}
\caption{The ratio $f$ versus the parameter $r_0$ for a fixed
$\xi=4$. Three curves from top to bottom correspond to the cases
of $\delta =0.3$, $0$ and $-0.3$, and corresponding
equation-of-state are $\omega_{e,0}= -1.47$, $-1.33$ and $-1.19$,
respectively.}
\end{figure}
In Fig.~12 we plot the ratio $f$ versus the scale $r_0$ for a
fixed $\xi$, but different $\delta$. It shows that a larger
$\delta$ gives a larger ratio for a fixed $r_0$. On the other
hand, we plot the ratio $f$ in Fig.~13 versus the scale $r_0$ for
a fixed $\delta$, but different $\xi$, which shows that a larger
$\xi$ gives us a larger ratio for a fixed $r_0$. Note that the
case with $\omega_{e,0}=-1.80$ is already ruled out~\cite{data}.
Here we draw this case just for illustration.  Fig.~14 shows the
relation of the ratio $f$  versus  the parameters $\xi$ and
$\delta$ for a fixed scale $r_0=5$.
\begin{figure}[ht]
\includegraphics[totalheight=1.7in]{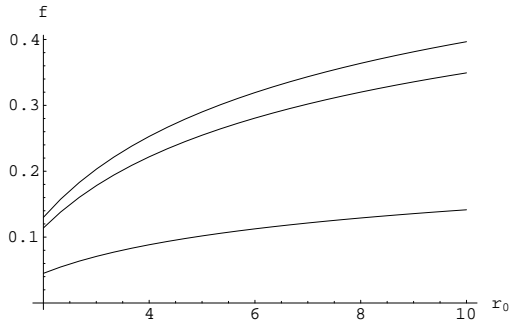}
\caption{The ratio $f$ versus the parameter $r_0$ for a fixed
$\delta=0.3$. Three curves from top to bottom correspond to the
cases of $\xi=5$, $4$ and $3$, and corresponding equation-of-state
 are $\omega_{e,0}= -1.80$, $-1.47$ and $-1.13$, respectively.}
\end{figure}
\begin{figure}[ht]
\includegraphics[totalheight=1.7in]{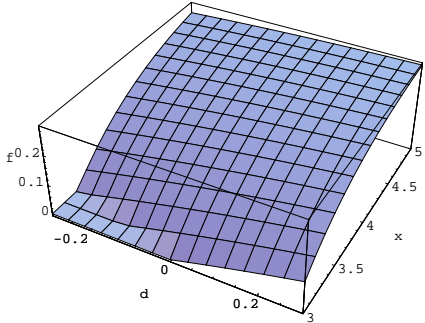}
\caption{The ratio $f$ versus the parameters  $\delta=d$  and
$\xi=x$ for a given $r_0=5$.}
\end{figure}

\sect{Conclusion}

In summary we discuss a cosmological model in which  phantom dark
energy has an interaction with dark matter. The interaction is
introduced phenomenologically by  considered an additional term
[see (\ref{2eq1}) and (\ref{2eq2})] in the equations of motion of
dark energy and dark matter. This term is parameterized by a
product of a dimensionless coupling function $\delta$,  Hubble
parameter and the energy density of dark matter, and it describes
an energy flow between the dark energy and dark matter. We discuss
two cases, one is the case where the equation-of-state
$\omega_e=p_e/\rho_e$ of the dark energy is kept as a constant;
the other corresponds to the case
 with a constant coupling function $\delta$.
  We investigate the effect of the interaction on the
evolution of the universe, the total lifetime of the universe, and
the fraction of the total lifetime of the universe for which the
universe is in the coincidence state, where the energy densities
of the dark energy and dark matter are comparable. We find that
the interaction has rich and significant consequences on these
issues. For example, the fraction of the total lifetime of the
universe for which the universe is in the coincidence state can
approximately reach $0.45$ if we take $\omega_e=-1.5$, $r_0=10$
and $\xi=3$. It means that the period when the energy densities of
the dark energy and dark matter are comparable is significantly
long, compared to its total lifetime.  Thus, it is not so strange
that we now live in the coincidence state of the universe. In this
sense the coincidence problem can indeed be significantly
ameliorated in such an interacting phantom dark energy model.
Finally let us stress that the constraints on the parameters of
the coupled dark energy model from the supernova Ia data has been
analyzed recently~\cite{Maje} (see also \cite{Dala,Amen}), the
values of parameters we take in this paper are all in the allowed
region. Certainly it is not enough to consider the constraints
from the supernova data only. We expect that the data from the
cosmic microwave background radiation and in particular, from the
large scale structure will give more restrictive constraints on
this coupled dark energy model. This issue is currently under
investigation.

{\bf Note added.} After the manuscript was put on the net, we have
been informed that a similar idea has also been considered by Z.K.
Guo and Y.Z. Zhang.  In addition, we have noticed that
following~\cite{Sch}, the coincidence problem has
 been discussed in a scalar field dark energy model with a linear effective
potential~\cite{Ave}.

\section*{Acknowledgments} RGC would like to express his gratitude
       to the Physics  Department, Baylor University for its warm hospitality.
       This work was supported by Baylor University, a grant from Chinese
       Academy of Sciences,  grants from NSFC (No. 10325525 and No. 90403029),
        and a grant from the Ministry of Science and Technology of China (No.
       TG1999075401).

\end{document}